\documentstyle[preprint,aps]{revtex}
\tighten
\begin{document}

\title{Bounds on $2m/R$ for static spherical objects}
\author{Jemal Guven${}^{(1)}$\thanks{e-mail \tt{ jemal@nuclecu.unam.mx}}
 and
   Niall \'O Murchadha${}^{(2)}$\thanks{e-mail \tt{ niall@ucc.ie}}}
\address{
${}^{(1)}$\
 Instituto de Ciencias Nucleares \\
 Universidad Nacional Aut\'onoma de M\'exico\\
 Apdo. Postal 70-543, 04510 M\'exico, D.F., MEXICO \\
${}^{(2)}$\ Physics Department,  University College Cork,\\
Cork, IRELAND \\
}
\maketitle
\begin{abstract}
It is well known that a spherically symmetric constant
density  static star, modeled as a perfect fluid, possesses a
bound on its mass $m$ by its radial size $R$ given by
$2m/R \le 8/9$ and that this bound continues to hold when the energy
density decreases monotonically.
The existence of such a bound is intriguing  because it occurs
well  before the appearance of an apparent horizon
at $m = R/2$. However, the assumptions made are extremely restrictive.
They do not hold in a simple soap bubble and they certainly do
not approximate any known topologically stable
field configuration. In addition, such configurations
will not generally be compact. We show that the 8/9 bound is robust by
relaxing these assumptions.
If the density is monotonically decreasing and the tangential stress is
less than or equal to the radial stress we show that the 8/9 bound
continues to hold through the entire bulk if $m$ is 
replaced by the quasi-local mass. If the tangential stress is allowed to 
exceed the radial stress and/or the density is not monotonic we cannot 
recover the 8/9 bound. However, we can show that
$2m/R$ remains strictly bounded away from unity by constructing an explicit
upper bound which depends only on the ratio of the stresses and the
variation of the density.

\end{abstract}
\date{\today}

\pacs{PACS numbers: 04.20.Cv}

\section{ INTRODUCTION}
Consider any static solution of the Einstein equations with the matter
satisfying the null energy condition. The Penrose singularity
theorem shows that this system cannot have a trapped surface.  
In a spherically symmetric configuration, the first apparent 
horizon in the  initial data occurs when the ratio of (twice) the 
mass to its radial extent, $2m/R$, is one. However, it is also well
known that for a spherical star composed of  ordinary matter with
positive energy described as a perfect fluid with a monotonically
decreasing energy profile,  $2m/R$ cannot exceed
8/9, the constant density value\cite{OV,Buch,Bondi,Wald,Wein}.  Such
a bound is particularly interesting because it occurs strictly before the
appearance of an apparent horizon. 
Here we wish to examine this bound more closely when the 
assumptions underlying it are relaxed.
Several very strong assumptions on the distribution of matter
enter its derivation. Even in an astrophysical  stellar object where
matter is described phenomenologically, it is not clear that the 
perfect fluid assumption is justified. A humble soap bubble
consisting of a membrane with a given (tangential) surface tension
supported by the pressure of the enclosed
perfect gas violates both the monotonicity of the energy density and the
perfect fluid assumption.  The approximation clearly also does not
represent accurately the interior of topological defects such as monopoles
\cite{VS}. The balance of forces providing 
the equilibrium typically turns out to be analogous to that
holding a soap bubble together. 
In addition, in many extensions of
Einstein gravity, the effective stress tensor describing a perfect fluid
does not assume the perfect fluid form.
Field configurations typically 
will not be compact, in which situation we require 
a generalization of the mass that
holds thoughout the bulk. This requires the replacement of 
$m$ by a quasi-local mass.

While the $8/9$ bound is not a universal one, it is robust in the sense
that  under physical conditions which are at least reasonable classically,
the mass continues to be bounded by a value strictly below the
apparent horizon value, $m= R/2$.
It appears to be impossible, even in principle, to construct a
static distribution which saturates it.

In Sections II and III, we establish our notation.
We show in Section III that if the matter satisfies $\rho + S_{\cal L} \ge
0$, where
$\rho$ is the energy density and $S_{\cal L}$ is the radial stress, the
object cannot have an apparent horizon and thus $2m/R$ is strictly bounded
away from 1. This condition, $\rho + S_{\cal L} \ge 0$ is one of the
so-called
`null energy' or `null convergence' conditions \cite{Hawking,morris,john}.
It is interesting because we need no restriction on the tangential stress
nor do we need to assume $\rho \ge 0$.
In a spherically symmetric geometry, the tangential stress will
generally differ from the radial one except at the center where the
constraints dictate that they coincide. Consider the ratio, $\gamma$,
of tangential to radial stress. In a perfect fluid $\gamma=1$.
We further show, again in Section III, that static matter
satisfying  $\rho \ge 0$
 together with $\gamma \le 1$
must have positive radial pressure which monotonically decreases outwards.
This guarantees that  $\rho + S_{\cal L} \ge
0$ which provides another way of excluding apparent horizons.

The rest of the article is devoted to investigating 
how close $2m/R$ can get to one, and generalizing the
 $2m/R < 8/9$ bound noted above. We 
summarize very briefly the simple constant density star in Section IV.
In Section V we consider `stars'  that are
monotonic with positive density. If
$\gamma\le 1$, it is simple
to show that the 8/9 inequality continues to hold, not only on the boundary
but through the entire bulk. Indeed, the configuration need not be compact. 
If $\gamma\ge 1$ anywhere, however, we obtain a slightly weaker result. We
construct a bound which shows that
$2m/R$ is strictly bounded away from unity. If $\gamma_{max}$ approaches 1
the bound smoothly approaches $8/9$; as $\gamma$ becomes unboundedly
large the bound approaches 1. In particular, for a monotonic star
with positive radial pressure and for which the transverse pressure
is less than the density we can show that $2m/R < 0.974$.

Finally, in Section VI, we relax
the assumption of monotonicity and find essentially the same results, except
that now the bound depends both on the variation of the matter as well as on
$\gamma_{max}$.

\section{Static Limit of Einstein Equations}

The spacetime metric
describing a static solution of the Einstein equations
can always be written in the form

\begin{equation}
ds^2 = - N^2 dt^2 + g_{ab} dx^a dx^b\,,
\label{eq:linel}
\end{equation}
where $N$ is the lapse function and the shift vanishes.
$N$ is also the norm of the global timelike
killing vector, $\partial_t$, and so must satisfy $N > 0$.
The spatial geometry at constant $t$ is described by the 
metric tensor $g_{ab}$. Both the material current
vector $J^a$ and the extrinsic curvature tensor $K_{ab}$ (describing the
embedding of a hypersurface of fixed $t$ in spacetime) vanish.
In the canonical formulation of the theory,
the momentum constraints of the theory are then  vacuous. The hamiltonian
constraint reduces to the form \cite{MTW,York}
(see also the appendix to \cite{Wald}),

\begin{equation}
{\cal R} = 16\pi \rho\,,
\label{eq:constr}
\end{equation}
where ${\cal R}$ is the scalar
curvature constructed with the spatial metric $g_{ab}$
and $\rho$ is the material energy density.
Given some specification of $\rho$, Eq.(\ref{eq:constr})
is a constraint on the spatial geometry, $g_{ab}$. It does not
involve the stresses operating on $\rho$.
In the spherically symmetric case we will see that the intrinsic geometry
is completely specified by $\rho$. The advantage of working within the
canonical formulation is that this constraint is isolated explicitly.

Given that the time direction is Killing the evolution in this
direction must be trivial. The dynamical Einstein equation reduces to
$\dot K_{ab}=0$, and now reads

\begin{equation}
-\nabla_a\nabla_b N + {\cal R}_{ab} N =
8\pi N \left(S_{ab} - {1\over 2} g_{ab} {\rm tr}\,S
+ {1\over 2} g_{ab}\rho\right)\,,
\label{eq:dynamic}
\end{equation}
where $\nabla_a$ is the covariant derivative compatible with
$g_{ab}$, ${\cal R}_{ab}$ is the associated Ricci tensor,
$S_{ab}$ is the material pressure tensor and $S$ is its trace.
In a perfect fluid the stress is isotropic with $S_{ab}= P g_{ab}$.
The other evolution equation, $\dot g_{ab} = 0$, is trivially satisfied.

For given $\rho$ and $g_{ab}$,
Eqs.(\ref{eq:dynamic}) consist of six PDEs for the seven
functions, $N$ and $S_{ab}$. This counting is not very precise, because
if we had a realistic fluid/field theoretical 
model we would have to supplement these
equations with an `equation of state' which would convert the equations
from an underdetermined to an overdetermined system.

We will suppose that the spatial topology is $R^3$.
For an object with
energy density of compact support (a star) or falling off sufficiently 
rapidly at infinity the spacetime will be asymptotically
flat with $N\to 1$ at infinity.
The appropriate boundary condition on $S_{ab}$ in 
an object of compact support is that its normal component
vanishes on the  surface.
For much of our discussion these boundary conditions are irrelevant.

Taking the trace of the equations, (\ref{eq:dynamic}),
and eliminating ${\cal R}$ in favor of $\rho$ using
Eq.(\ref{eq:constr}),
we obtain the linear elliptic equation for $N$,

\begin{equation}
\Delta N = 4\pi (\rho + {\rm tr}\,S) N \,.
\label{eq:trace}
\end{equation}
If the strong energy condition is satisfied we have $\rho + {\rm tr}\,S\ge
0$  everywhere, and so $\Delta N \ge 0$ when $N>0$. Thus the solution
cannot possess an interior maximum.
$N$ falls towards the center. Even if the potential, $V= \rho + {\rm
tr}\,S$, is large, $N$ never fall to zero in the interior
(e.g.,\cite{NOM}). The lapse can go to zero only when the density or
pressure becomes unboundedly large.

Let us assume that $N$ is positive in the exterior and negative on a
compact region $W$. $N$ vanishes on $\partial W$ but will have positive
outward gradient. If we integrate $\Delta N$ over $W$ we can turn it into a
surface integral which must be positive. On the other hand, from
Eq.(\ref{eq:trace}) we see that $\Delta N \le 0$ on $W$, so we have a
contradiction. We also see that it is impossible for $N$ to just touch zero
at a point. At that point we would have that $N$, its first derivatives and
its second derivatives all vanish. Thus the function could never grow
away from zero\cite{murray}.

One can deduce the conservation law,

\begin{equation}
 \nabla_b S^{ab} = - (S^{ab} + \rho g^{ab} ) {\nabla_b N\over N} \,,
\label{eq:conserv}
\end{equation}
directly from the static Einstein equations,
Eqs.(\ref{eq:constr}) and (\ref{eq:dynamic}). To do this,
we simply take the divergence of Eq.(\ref{eq:dynamic}). Exploiting
the Ricci identities,

\begin{equation}
[\nabla_a, \nabla_b] V^b = R_{ab} V^b \,,
\end{equation}
and the contracted Bianchi identity for ${\cal R}_{ab}$,
$\nabla_a {\cal R}^{ab} = \nabla^b {\cal R}/2$, we reproduce
Eq.(\ref{eq:conserv}).

It is clear from Eq.(\ref{eq:trace}) that there are no
non-trivial vacuum  static solutions in the theory.
We have $\Delta N=0$ everywhere. If there is
no internal boundary, the  solution is $N=1$ everywhere.
Now ${\cal R}_{ab}=0$, as well as ${\cal R}=0$,
so that the geometry is flat everywhere.
There is a well known result that
the only perfect fluid static equilibria are
spherically symmetric \cite{Masood,Beig}.
This result implies that for perfect fluids the spherically
symmetric analysis is complete.
A discussion of the symmetries of 
equilibrium configurations is provided in \cite{Lind}.

\section{Spherical Symmetry}

The line element describing the spatial part of a spherically symmetric
geometry  can always be written as

\begin{equation}ds^2= d\ell^2+R^2 d\Omega^2\,,\label{eq:lineel}\end{equation}
$\ell$ is the proper radial distance on the hypersurface,
$R$ is the areal radius. For $R^3$ topology, $\ell$ has domain $[0,\infty)$.
The appropriate boundary conditions on $R$ are

\begin{equation}
R(0) =0\,, \hskip 1cm dR/d\ell|_0 = R'(0) =1\,.
\end{equation}
The scalar curvature ${\cal R}$ is given by

\begin{equation}
{\cal R} = -{2\over R^2}\Big[2 \left(R R^\prime \right)^\prime -R^{\prime 2}
-1 \Big]\,,\label{eq:ricci}\end{equation}
where primes denote derivatives with respect to $\ell$.
The constraint equation can be cast in the form
(see, for example, \cite{I,II})

\begin{equation}
R'^2 = 1 - {2 m\over R}\,,
\label{eq:R'2}
\end{equation}
where the positive quasi-local mass is given by

\begin{equation}
m = 4\pi \int_0^\ell \rho R^2 R' d\ell = 4\pi \int_0^R \rho R^2 dR\,.
\label{eq:qm}
\end{equation}
It is immediately clear that

\begin{equation}
m\le R/2
\label{eq:hor}
\end{equation}
everywhere.
In general, $R'^2\le 1$ in any regular geometry when the
weak energy condition $(\rho \ge 0)$ is satisfied, so that
$R\le \ell$ everywhere\cite{II}. To show this we substitute
Eq.(\ref{eq:ricci}) into Eq.(\ref{eq:constr}) to get
\begin{equation}
2RR'' + R'^2 - 1 = - 8\pi R^2\rho\,.
\label{eq:con}
\end{equation}
At the center we have $R' = 1$ and $m = 0$. From Eq.(\ref{eq:qm}) we see
that $m$ increases as soon as we meet matter and thus $R'$ drops below
1. Let us assume that it later rises up to $+1$. However, from
Eq.(\ref{eq:con}) we see that at this point $R'' \le 0$ so it cannot be
rising! On the other hand $R'$ can drop below $-1$. 
We again get $R'' \le 0$
which means that it cannot ever rise up again to the asymptotic $R'\approx
1$. Thus for any regular spherical geometry satisfying the weak energy
condition $-1 < R' \le +1$ and $R' = +1$ only at the origin and at infinity.
This holds true for any solution of Eq.(\ref{eq:con}), no static assumption
is required.

It is clear from Eq.(\ref{eq:R'2}) that
$m$ is positive everywhere in a regular geometry and vanishes only at the
center and in any vacuum region surrounding it. In a static configuration,
the extrinsic curvature vanishes so that an apparent horizon is a minimal
surface with 
$R'=0$. Thus, if the geometry is free of an apparent horizon, it
must also be free of singularities. In such a geometry $0 < R' \le 1$ and
$m$ increases monotonically with $\ell$ (or $R$).
We emphasise  that the spatial geometry and with it the ADM mass
is completely determined by the source energy density.
The material stresses play no role whatever.

At the surface of a compact object of radius $R=R_0$, the quasi-lcoal
mass coincides with the constant ADM mass, $m_0$. 
The exterior solution is given by Eq.(\ref{eq:R'2})

\begin{equation}
R'^2 = 1 - {2 m_0\over R}\,.
\label{eq:Sch}
\end{equation}

In a spherically symmetric geometry,
any symmetric tensor is completely
characterized by two scalars. We have

\begin{eqnarray}
{\cal R}_{ab} &=& {\cal R}_{\cal L} n_a n_b + {\cal R}_R (g_{ab} - n_a n_b)
\nonumber\\
S_{ab} &=& S_{\cal L} n_a n_b  + S_R (g_{ab} - n_a n_b)\,.
\end{eqnarray}
Here $n^a$ is the outward pointing normal to a two sphere of fixed
proper radius. The two scalars appearing in the Ricci tensor can be
expressed in terms of ${\cal R}$, $R$ and $R'$ as follows \cite{II}:

\begin{eqnarray}
{\cal R}_{\cal L} &=& {1\over 2}{\cal R} - {1\over R^2} (1 - R'^2)\nonumber\\
{\cal R}_R &=& {1\over 4}{\cal R} + {1\over 2R^2} (1 - R'^2)\,.
\end{eqnarray}
Taking the two independent projections
of Eq.(\ref{eq:dynamic}), we therefore have in any spherically symmetric
static equilibrium,
\begin{eqnarray}
N'' &=& N
\left\{ 4\pi (\rho - S_{\cal L} + 2 S_R ) - {2m\over R^3}
\right\}\label{eq:E1}\\
R' N' &=& R N
\left\{ 4\pi S_{\cal L}  + { m\over R^3}
\right\}\,.
\label{eq:E2}
\end{eqnarray}
We can also combine Eqs.(\ref{eq:E2}) and (\ref{eq:con}) to obtain
\begin{equation}
-{R'' \over R} + {R'N' \over RN} = 4\pi(\rho + S_{\cal L})\,.
\label{eq:E3}
\end{equation}
These three equations are the complete set of equations satisfied by any
static spherically symmetric system.

If the matter satisfies
\begin{equation}
\rho + S_{\cal L} \ge 0\,,
\label{eq:se}
\end{equation}
we can immediately deduce from Eq.(\ref{eq:E3}) that any spherical
static configuration cannot have an apparent horizon. 
The apparent horizon coincides with a minimal surface,
i.e., when $R' = 0$. From (\ref{eq:E3}) at a minimal surface we see that
$R'' \le 0$. However, at the outermost minimal surface we must have $R'' >
0$ since the area is increasing outwards. Thus we have a contradiction.
Thus we have shown that, in a static star satisfying Eq.(\ref{eq:se}), $R'
> 0$. From Eq.(\ref{eq:R'2}) this is equivalent to showing $m < R/2$.
In \cite{morris} essentially the converse of this argument was given,
showing that if a minimal surface existed (the throat of a static
wormhole) then the matter cannot satisfy Eq.(\ref{eq:se}).
This result has been recently proven in
\cite{Rend}.  However, that proof requires that both $\rho$ and $S_{\cal L}$
be positive while we only need to impose a condition on the combination.

The energy condition Eq.(\ref{eq:se}) is a single component of what is
called the `null convergence' or `null energy' condition
 \cite{Hawking,john}. It is a consequence of each of the three standard
energy conditions (the `strong', `weak', and `dominant').  Consider the
outgoing radial null vector
\begin{equation}
\xi^{\mu} = (1/N, 1, 0, 0)\,,
\end{equation}
and multiply it into the spacetime Ricci tensor 
$^{(4)}\cal{R}_{\mu \nu}$ to get
\begin{equation}
^{(4)}\cal{R}_{\mu \nu}\xi^{\mu}\xi^{\nu} =  ^{(4)}\cal{G}_{\mu
\nu}\xi^{\mu}\xi^{\nu} = {\rm 8} \pi (\rho + S_{\cal
L})\,.
\label{eq:tc}
\end{equation}
where  
$^{(4)}\cal{G}_{\mu \nu}$ is the spacetime
Einstein tensor. The equality above is to be expected because the
Einstein tensor only differs from the Ricci tensor by a trace and the trace
term, when dotted twice with a null vector, vanishes.
The positivity of $^{(4)}\cal{R}_{\mu \nu}\xi^{\mu}\xi^{\nu}$ 
implies Eq.(\ref{eq:se}).
Choosing $\xi^\mu$ to be an outgoing tangential null vector we obtain
$\rho + S_{\cal L} + 2S_R \ge 0$.

If both $S_{\cal L}\ge 0$ and  $\rho \ge 0$
hold independently, as supposed in \cite{Rend},
it is clear that Eq.(\ref{eq:se})
is satisfied which guarantees
$R' > 0$. We also now have that the right hand side of Eq.(\ref{eq:E2}) is
positive so that $N'
\ge 0$ everywhere.  The lapse function, the length of the Killing vector,
for any  regular solution must grow monotonically out from the centre to its
asymptotic value one. With positive radial stress and positive $\rho$ we do
not need to assume the strong energy condition.
Spherical symmetry is very restrictive.
Compare this to the spherically symmetric statement of the maximum principle
which was applied earlier to the trace equation, Eq.(\ref{eq:trace}).

For completeness,
we note that the lapse is evaluated in the exterior of a compact object
as follows: using Eq.(\ref{eq:E2}), we have

\begin{equation}
R' N' =  N { m_0 \over R^2}\,,
\end{equation}
so that using Eq.(\ref{eq:Sch}),

\begin{equation}
{N'\over R'} =
{ m_0 \over R^2}\left( 1- {2 m_0\over R}\right)^{-1}\,.
\end{equation}
The boundary condition at infinity, $N\to 1$, fixes

\begin{equation}
N = \left( 1- {2 m_0\over R}\right)^{1/2}\,.\label{eq:extlap}
\end{equation}
Eq.(\ref{eq:extlap}) together with
Eq.(\ref{eq:Sch}) reproduce the exterior Schwarzschild form of the
spacetime metric.

The conservation of the stress tensor reduces to the
single equation,

\begin{equation}
S_{\cal L}' + 2 {R'\over R} (S_{\cal L} - S_R) = - ( S_{\cal L} + \rho)
{N'\over N} \,.\label{eq:SN}
\end{equation}
We deduce immediately that at $\ell=0$ in a non-singular geometry

\begin{equation}
S_{\cal L} = S_R\,.
\end{equation}
The perfect fluid form of the stress tensor is the only
one consistent with the symmetry at the origin. This is exactly
as in newtonian theory.

While we only needed a condition on the radial stress to eliminate apparent
horizons, the transverse stress does play a role in 
establishing the equilibrium. 
This can be seen in simple mechanical models. For example, in a
soap bubble, the surface tension, which is effectively a negative
transverse stress, is the object which balances the positive outward
pressure difference between the inside and outside. On the other hand, 
if we had an evacuated spherical metal shell, with a vacuum 
inside and positive pressure
outside, the outside pressure forces the metal shell to contract 
setting up a positive transverse stress. The radial stress obviously
increases outwards and is balanced by the positive transverse stress. In a
self-gravitating system, we expect the radial pressure to decrease
outwards.  However, it need not if there
are large positive transverse stresses to support the external pressure.

In a spherically symmetric geometry it is possible to
exploit the first order Einstein equation,
Eq.(\ref{eq:E2}), to reduce the dependence
on the stress tensor appearing in Eq(\ref{eq:E1})
to a dependence on the ratio of the tangential to the radial stress,

\begin{equation}
\gamma = {S_R\over S_{\cal L}}\,.
\end{equation}
We have
\begin{equation}
N'' + {R'\over R} \left(1 - 2\gamma\right) N' =
\left\{ 4\pi \rho - {m\over R^3} \left(1 + 2\gamma \right) \right\}
N\,.
\label{eq:Ngamma}
\end{equation}
In particular, if the stress is isotropic then $\gamma=1$ and
Eq.(\ref{eq:Ngamma}) is independent of $S_{ab}$.

Alternatively, we can exploit Eq.(\ref{eq:E2})
to eliminate the lapse from the
conservation equation, Eq.(\ref{eq:SN})

\begin{equation}
S_{\cal L}' + 2 {R'\over R} (S_{\cal L} - S_R) = - {R\over R'}
( S_{\cal L} + \rho)
\left( 4\pi S_{\cal L}  + { m\over R^3}
\right)\,.
\label{eq:SS}
\end{equation}
In the isotropic limit, Eq.(\ref{eq:SS})  is the
Tolman-Oppenheimer-Volkov equation.
In the newtonian limit, the RHS of Eq.(\ref{eq:SS}) is
replaced by $- \rho m / r^2$.

In the simple mechanical models provided above, it is clear that positive
transverse stresses reduce the internal pressure and {\it vice versa}. We
can exploit Eq.(\ref{eq:SN}) (or Eq.(\ref{eq:SS})) to show that, in
general, if the transverse pressure is smaller than the radial pressure the
radial pressure builds up inside. We will give two slightly different
versions of this result.

First, let us assume  $S_{\cal L} - S_R \ge 0$, 
$\rho + S_{\cal L} \ge 0$
and
$\rho + S_{\cal L} + 2S_R \ge 0$ (both of the latter
coming from the `strong energy'
condition). From the trace equation,
Eq.(\ref{eq:trace}), we have  that $N' > 0$ and from Eq.(\ref{eq:E3}) we
have $R' > 0$. When these are substituted into  Eq.(\ref{eq:SN}) we get
$S'_{\cal L} < 0$ so the pressure monotonically increases inward.
The object  does not need to be compact.

Alternatively, even more simply, let us assume  $S_{\cal L} - S_R \ge 0$ and
$\rho \ge 0$. The
hamiltonian constraint guarantees $m >0$. Suppose first that the 
object is compact. We cannot
have an apparent horizon outside so we have $R' > 0$ on the
boundary. At the  boundary of a compact object 
$S_{\cal L} =0$ and so from Eq.(\ref{eq:SS}) we get
$S_{\cal L}'<0$ so that
$S_{\cal L}$ is decreasing outwards and so must be positive near the
boundary. However a positive $S_{\cal L}$ makes the right hand
side of Eq.(\ref{eq:SS}) even more negative so $S_{\cal L}$ becomes ever
larger as one moves inwards. Thus we have shown that $S_{\cal L} > 0$ and
monotonically decreases as one travels out. In turn this guarantees
both $\rho + S_{\cal L} > 0$ and $4 \pi S_{\cal L} + m/R^3 >0$. Hence
$R' > 0$ and
$N' > 0$. Note that we need not assume that $S_R$ vanishes on the boundary
but it cannot be positive there. In other words, surface tension is good.
If the object is nor compact, the argument we have just presented
is valid in the region bounded by any sphere with $S_{\cal L}\ge 0$.

In this section, using very weak assumptions, we have demonstrated that in
a spherical static star we have $0 < R' \le 1$. From Eq.(\ref{eq:R'2}) we
now get that both $m > 0$ and $2m/R < 1$. However, we have to date no
information  on how close $R'$ can get to zero, how close $2m/R$ can get to
1. This will be discussed in the next sections, where, by imposing various
restrictions on the matter, we get extra control on the behaviour of $2m/R$.

\section{Constant Density Perfect Fluid Star}

It is our good fortune
that for a perfect fluid constant density star,
Eq.(\ref{eq:SS}) is exactly solvable.
Eqs.(\ref{eq:R'2}) and (\ref{eq:qm})
reduce to

\begin{equation}
R'^2 + \left({8\pi\rho_0\over 3}\right) R^2 = 1 \,.
\label{eq:cds}
\end{equation}
We then have

\begin{equation}
{d P\over dR}  = -  4\pi {R\over 1- {8\pi\over 3} \rho_0 R^2}
( P + \rho_0)
\left( P   + { 1\over 3}\rho_0 \right)\,,
\label{eq:PP}
\end{equation}
with the well known solution,

\begin{equation}
\label{P}
P (R) = \rho_0 \,\left({
\left(1 - {2m_0 R^2\over R_0^3}\right)^{1/2}
- \left(1 - {2m_0\over R_0}\right)^{1/2} 
\over
3 \left(1 - {2m_0\over R_0}\right)^{1/2} 
- \left(1 - {2m_0 R^2 \over R_0^3}\right)^{1/2}}
\right)\,.
\end{equation}
The pressure  always exceeds the newtonian value. In fact,
in an isotropic uniform newtonian fluid ball of radius $R_0$,
the central pressure is given by $P=P_c$, where

\begin{equation}
P_c = {2\pi\over 3} \rho_0^2 R_0^2\,.
\end{equation}
The pressure given by Eq.(\ref{P}) diverges at the center $R=0$
when
$m_0 = 4 R_0/9$. This occurs when the surface lapse, $N_0=1/3$.
If $m_0> 4R_0/9$, it diverges at some finite value of $R$.
As $2m_0$ is increased up to $R_0$, the divergence moves out to $R_0$.

Let us now examine the lapse. In a constant density perfect fluid,
Eq.(\ref{eq:Ngamma}) assumes the very simple form,

\begin{equation}
\left({N'\over R}\right)' = 0\,.
\end{equation}
We exploit the continuity of the
lapse and its first derivative across $R_0$ which follow from
Eq.(\ref{eq:E1}) and (\ref{eq:E2})
We first integrate out from some interior point to
the surface at $R_0=R(\ell_0)$:

\begin{equation}
\left({N'\over R}\right)_{R<R_0} =
\left({N'\over R}\right)_{R=R_0} = {m_0\over R_0^3}\,,
\end{equation}
where we have exploited Eq.(\ref{eq:extlap}) to evaluate the
RHS. We integrate again over the same domain.
We find for the surface lapse,

\begin{equation}
N_0 =  N_c +
{m_0\over
R_0^3}\int_0^{\ell_0} d\ell \,R(\ell) \,,
\label{eq:N(0)}
\end{equation}
where $N_c$ is the value of the lapse at the center.
We note that generally

\begin{equation}
\int_\ell^{\ell_0} d\ell R(\ell) =
\int_R^{R_0} R \,dR
\left(1 - {2m\over R} \right)^{-1/2} \,.
\label{eq:int}
\end{equation}
Thus, in a constant density star,

\begin{equation}
\int_\ell^{\ell_0} d\ell R(\ell) =
\int_R^{R_0} R\, dR
\left(1 - {2m_0R^2\over R_0^3} \right)^{-1/2} =
{1\over 2}\left( 1 -
\left( 1- {2 m_0 R^2\over R_0^3}\right)^{1/2}
\right)
{R_0^3\over m_0} \,,
\end{equation}
and we get
\begin{equation}
N_0 = \left( 1 - {2m_0\over R_0}\right)^{1/2}  = N_c +  {1\over
2}\left( 1 -
\left( 1- {2 m_0 \over R_0}\right)^{1/2}
\right)\,.
\label{eq:bdd}
\end{equation}
We require $N_c \ge 0$. 
Eq.(\ref{eq:bdd}) then implies
\begin{equation}
0\le
{3\over 2}\left( 1- {2 m_0\over R_0}\right)^{1/2}
- {1\over 2}\,,
\label{eq:4/9}
\end{equation}
or

\begin{equation}
m_0 \le {4\over 9} \,R_0\,,
\end{equation}
exactly as before.
This route, however, has the advantage that Eq.(\ref{eq:Ngamma}) is
linear in $N$ unlike Eq.(\ref{eq:SS}) which is nonlinear in $S_{\cal L}$.

It is worth noting that $N \rightarrow 0$ as $m \rightarrow 4R/9$ should
not be viewed as the Killing vector going null. It is another version of
the `collapse of the lapse' phenomenon, in this case driven by the fact
that the pressure is becoming unboundedly large.

\section{ Monotonic Stars}

Buchdahl \cite{Buch} demonstrated that if the energy density profile in a
star is monotonically decreasing, and it is modeled
as a perfect fluid, this 4/9 bound continues to hold. The
constant density star saturates the bound within this class of
systems. In this section we follow Buchdahl in only considering objects with
monotonically decreasing densities but we will push the calculations much
further. We start off with a perfect fluid assumption and rederive the 4/9
bound. We then weaken this to the dominant radial pressure assumption
$(S_{\cal L} \ge S_R)$ that we used in Section III and prove that the 4/9
bound is still valid. 
We next extend the inequality to 
interior points. We finally consider the situation where $S_R$ may be
larger than $S_{\cal L}$. We no longer can recover the 4/9 bound; however
if the ratio of the pressures is bounded we show that $m/R$ is strictly
bounded away from 1/2.

Let us define
\begin{equation}
{4\pi\over 3} 
\langle\rho\rangle :=
{m\over R^3}\,,
\end{equation}
so that

\begin{equation}
\langle\rho\rangle = {\int_0^R \rho R^2 dR\over
\int_0^R R^2 dR}\,,
\end{equation}
is an average of $\rho(R)$ 
(not to be confused with the physical average)
within a euclidean ball.
Thus if $\rho'\le 0$, it is clear that $\langle\rho\rangle \ge \rho$ and

\begin{equation}
\left({m\over R^3}\right)' = {4\pi\over 3} \langle\rho\rangle' \le 0\,.
\end{equation}
In particular, one can deduce that  $m/R\ge m_0 R^2/R_0^3$, so that
\begin{equation}
\left(1 - {2m\over R} \right)^{-1/2}\ge
\left(1 - {2m_0R^2\over R_0^3} \right)^{-1/2}\,,
\label{eq:lb}
\end{equation}
 a lower
bound is always provided by a constant density star with the
same $m_0$ and $R_0$.

We mimic the constant density star calculation.
This is essentially the Buchdahl derivation, however we allow for a
non-perfect fluid.  We combine Eqs.(\ref{eq:E1}) and (\ref{eq:E2}) to give
\begin{equation}
\left({N'\over R}\right)' = {N'' \over R} - {N'R' \over R^2} = {4\pi N \over
R}
\left[\left(\rho - \langle\rho\rangle\right) + 2\left(S_R - S_{\cal
L}\right)\right]
\label{eq:key}
\end{equation}
Both terms on the RHS of Eq.(\ref{eq:key})
are negative when $\rho'\le 0$ and $S_{\cal L} - S_R \ge 0$.
Thus we have

\begin{equation}
\label{N'}
\left({N'\over R}\right)' \le 0\,,
\end{equation}
with equality only in a constant density star supported by
isotropic pressure.
The remainder of the calculation in this case mimics that
for a constant density star.

As before, we first integrate Eq.(\ref{N'}) 
out from some interior point to
the surface at $R=R_0$:

\begin{equation}
\label{N1}
\left({N'\over R}\right)_0 \ge
{m_0\over R_0^3}\,.
\end{equation}
We follow this by  integrating
out from the center at $\ell=0$ to the surface. We find

\begin{equation}
N_0 \ge N_c +
{m_0\over R_0^3}\int_0^{\ell_0} d\ell\, R(\ell) \,.
\label{eq:N1(0)}
\end{equation}
We require a lower bound on the integral appearing in the second term on the
RHS. We cast it, as before, in the
form (\ref{eq:int}).
Using Eq.(\ref{eq:lb}), it is clear that

\begin{equation}
\int_0^{\ell_0} d\ell\, R(\ell) \ge
{1\over 2}\left( 1 -  \left( 1- {2 m_0\over R_0}\right)^{1/2}\right)
{R_0^3\over m_0} \,.\label{*}
\end{equation}
When we substitute Eq.(\ref{*}) and Eq.(\ref{eq:extlap}) into
(\ref{eq:N1(0)}) together with the  requirement that $N_c\ge 0$
we recover Eq.(\ref{eq:4/9}) and so we have that $2m_0/R_0 \le 8/9$.

This gives only a bound at the boundary. If the configuration has a
`thin' atmosphere with $3\rho < \langle\rho\rangle$, $m/R$ is decreasing so
the maximum value of $m/R$ occurs
somewhere in the interior and not on the
boundary.  In such a scenario the above result is not very useful.
Happily, the argument can be tweaked to show that $2m/R \le
8/9$ through the whole system.

Let us assume $\rho \ge 0$, $\rho' \le 0$ and $S_{\cal{L}} \ge S_R$. The
argument at the end of Section III shows us that  $S_{\cal{L}} \ge 0$.
From Eq.(\ref{eq:E2}) 
we have
\begin{equation}
 {N'\over R} = {N \over R'}
\left( 4\pi S_{\cal L}  + { m\over R^3}
\right) \ge {N \over R'}
\left({ m\over R^3}
\right)\,.
\label{**}
\end{equation}
Let us assume that $2m/R$ possesses a maximum 
at a point a distance $\ell_1$ from the
center. The monotonicity of $N'/R$ (Eq.(\ref{N'})) and
Eq.(\ref{**}) gives
(contrast Eq.(\ref{N1}))

\begin{equation}
 {N'\over R} \ge \left({N \over R'}
{ m\over R^3}
\right)_1 \forall\; \ell \le \ell_1\,.
\label{***}
\end{equation}
Integrate from the center to $\ell_1$ to get

\begin{eqnarray}
N_1 &\ge & N_c + \left({N \over R'}{ m\over R^3}\right)_1
\int_0^{\ell_1}R\, d\ell \\
& = & N_c + \left({N \over R'}{ m\over R^3}\right)_1
\int_0^{R_1}{R dR \over \left( 1 - 2m/R\right)^{1/2}} \\
& \ge &  N_c + \left({N \over R'}{ m\over R^3}\right)_1
\int_0^{R_1}{R dR \over \left(1 - {2m_1R^2 /R_1^3}\right)^{1/2}}
\label{eq:N_1}\,,
\end{eqnarray}
where the last line
follows from the monotonicity of $m/R^3$. 
This can be integrated to give
\begin{equation}
N_1 \ge N_c + {m_1N_1 \over R_1^3\left(1 - {2m_1\over R_1}\right)^{1/2}}
{R_1^3 \over 2m_1}\left[1 - \left(1 - {2m_1\over R_1}\right)^{1/2}\right]\,.
\end{equation}
Requiring that $N_c \ge 0$ allows us to cancel the $N_1$ on both sides and
we immediately get that $2m/R \le 8/9$.

To deal with the situation where $S_{\cal{L}}$ can be less than $S_R$ we
need a somewhat more complicated argument. We add now as one of our
assumptions that $S_{\cal L} \ge 0$.

If we divide Eq.(\ref{eq:key}) by
Eq.(\ref{eq:E2}) we can get

\begin{equation}
\left({N'\over R}\right)'  = { N' \over R}{R' \over R}
{\left(\rho - \langle\rho\rangle\right) + 2\left(S_R - S_{\cal
L}\right) \over S_{\cal{L}} + {\langle\rho\rangle \over 3}}\,.
\label{eq:key1}
\end{equation}
Let us assume that the term on the right hand side of Eq.(\ref{eq:key1})
which depends on the sources is bounded. In other words we assume
\begin{equation}
{\left(\rho - \langle\rho\rangle\right) + 2\left(S_R - S_{\cal
L}\right) \over S_{\cal{L}} + {\langle\rho\rangle \over 3}} \le \beta \,.
\label{eq:key2}
\end{equation}
It is clear that $\beta$ cannot be negative because the numerator vanishes
at the center. We have $\beta = 0$ for a monotonic star with $S_R \le
S_{\cal L}$. In general, it will be some positive number.
Eq.(\ref{eq:key1}) now reads
\begin{equation}
\left({N'\over R}\right)'  \le \beta { N' \over R}{R' \over R}\,.
\label{eq:key3}\end{equation}
Find the point where $2m/R$ is a maximum (call it $\ell_1$ as before) and
integrate Eq.(\ref{eq:key3})  out to it to give
\begin{equation}
\ln \left({(N'/R)_1\over (N'/R)}\right) \le
\beta \ln R_1/R \,,
\end{equation}
so that
\begin{equation}
{N'\over R}\ge \left({N'\over R}\right)_1
\left({R\over R_1}\right)^{\beta}\,.
\end{equation}
As before, we integrate this equation from the center out to $\ell_1$ to get
\begin{eqnarray}
N_1 &\ge & N_c + \left({N' \over R}\right)_1
\int_0^{\ell_1}\left({R \over R_1}\right)^{\beta}R d\ell\label{N_5} \\
N_1 &\ge & N_c + \left({N \over R'}\left[S_{\cal L} + {m\over
R^3}\right]\right)_1
\int_0^{\ell_1}\left({R \over R_1}\right)^{\beta}R d\ell\label{N_2} \\
& \ge & N_c + \left({N \over R'}{ m\over R^3}\right)_1
\int_0^{R_1}\left({R \over R_1}\right)^{\beta}{R dR \over 
\left( 1 - 2m/R\right)^{1/2}}
\label{N_3}\\
& \ge &  N_c + \left({N \over R'}{ m\over R^3}\right)_1
\int_0^{R_1}\left({R \over R_1}\right)^{\beta}{R dR \over \left(1 -
{2m_1R^2 \over R_1^3}\right)^{1/2}}
\label{eq:N_4}\,.
\end{eqnarray}
In going from Eq.(\ref{N_5}) to Eq.(\ref{N_2}) we use
Eq.(\ref{eq:E2}) and
in going from Eq.(\ref{N_2}) to Eq.(\ref{N_3}) we  use that
$S_{\cal L}(\ell_1) \ge 0$. It is clear that the integral in Eq.(\ref{eq:N_4})
is finite and well behaved for any finite $\beta$. Thus we get a bound
on $2m/R$ which is strictly bounded away from 1. Only in the limit as
$\beta \rightarrow \infty$ does the integral go to zero. In this case the
bound on
$2m/R \rightarrow 1$. In the other limit, when $\beta \rightarrow 0$, we
recover Eq.(\ref{eq:N_1}) and so we get $2m/R \rightarrow 8/9$. In the
special cases where $\beta = 2, 4, 6, \dots$ the integral in
Eq.(\ref{eq:N_4}) can be done simply. This includes one
especially interesting case.

Let us assume we are given a monotonic star with positive radial pressure
(these assumptions can be justified by stability criteria).
Let us further assume that the transverse pressure is bounded. More
precisely let us assume $S_R \le \rho$. This can be justified on some kind
of speed of sound argument. From the monotonicity we get $S_R \le \langle
\rho \rangle$. From these we immediately get $\beta \le 6$! Now we can do
the integration and get $2m/R \le 0.974$.

Alternatively, if the material was approximately a perfect
fluid we could use the ratio of the pressures, $\gamma$, that we introduced
earlier. It is clear that $\beta \le 2(\gamma_{Max} - 1)$.

\section{Completely general spherical configuration}

Let us now consider a general static spherical ball. We no longer wish to
assume either monotonicity or a perfect fluid. The only constraints we
place are that both $\rho \ge 0$ and $S_{\cal L} \ge 0$. We also assume
that $\beta$ as defined by Eq.(\ref{eq:key2}) exists. We are less interested
in obtaining the tightest bound on
$2m/R$ than in  establishing that such a bound exists.

All the equations, starting from  Eq.(\ref{eq:key2} up to and including
Eq.(\ref{N_3}) continue to hold. However, in going from Eq.(\ref{N_3}) to
Eq.(\ref{eq:N_4}) we used the monotonicity. One way of avoiding that
difficulty is by replacing Eq.(\ref{eq:N_4}) with
\begin{equation}
 N_1 \ge   N_c + \left({N \over R'}{ m\over R^3}\right)_1
\int_0^{R_1}\left({R \over R_1}\right)^{\beta}{R dR \over \left(1 -
{8\pi\langle \rho\rangle_{\rm Min}\over 3}R^2\right)^{1/2}}
\label{eq:N_6}\,.
\end{equation}
This uses that  $2m/R= 8\pi \langle \rho\rangle
R^2/3 \ge  8\pi \langle \rho\rangle_{\rm Min}
R^2/3$.
Eq.(\ref{eq:N_6}) can be simplified by introducing a new variable $x^2 =
8\pi \langle \rho\rangle_{\rm Min}R^2/3$ and $x^2_1 =
8\pi \langle \rho\rangle_{\rm Min}R^2_1/3 = 2\bar{m}_1/R_1 \le 2m_1/R_1$
where $\bar{m} =
4\pi \langle \rho\rangle_{\rm Min}R^3/3 \le m$. We then get from $N_c > 0$

\begin{equation}
\sqrt{1 - {2m_1 \over R_1}} \ge {\langle \rho\rangle \over 2\langle
\rho\rangle_{\rm Min}}\int_0^{x_1}\left({x \over x_1}\right)^{\beta}{x dx
\over \sqrt{1 - x^2}}\label{N_7}\,.
\end{equation}
It is clear that the right hand side of Eq.(\ref{N_7}) is finite and
bounded away from zero as long as $\langle \rho\rangle_{\rm Min}$ is
non-zero. This is very misleading because of the dependence of $x_1$ on
$\langle \rho\rangle_{\rm Min}$. If we return to Eq.(\ref{eq:N_6}) we can
see that the integral has a lower bound of $R^2_1/(\beta + 2)$ and this is
achieved when  $\langle \rho\rangle_{\rm Min} = 0$. In this case we get a
bound on $2m_1/R_1$ given by
\begin{equation}
2(\beta + 2)\sqrt{1 - {2m_1 \over R_1}} \ge {2m_1 \over R_1}\,.
\end{equation}
Thus we get the following bound
\begin{equation}
{2m_1 \over R_1} \le {1\over 2}(\beta + 2)^2
\left[\sqrt{1 + {4 \over (\beta + 2)^2}}
- 1 \right] \approx 1 - {2 \over (\beta + 2)^2}\,.\label{B_1}
\end{equation}
The approximation given in Eq.(\ref{B_1}) holds only in the limit as
$\beta$ becomes large. In general, we can show that the expression in
Eq.(\ref{B_1}) is always less than 1 and monotonically increases with
$\beta$.
For example, when $\beta = 0$ we get $2m_1/R_1 \le 0.944$.

 We know that
$0 \le \langle \rho\rangle_{\rm Min} \le \langle \rho\rangle_1$. For any
fixed value of $\beta$, if $\langle \rho\rangle_{\rm Min} \approx \langle
\rho\rangle_1$ the bound we get on $2m_1/R_1$ agrees with the monotonic
bound as given implicitly by Eq.(\ref{eq:N_4}). As $\langle
\rho\rangle_{\rm Min}$ reduces below $\langle \rho\rangle_1$ the bound on
$2m_1/R_1$ increases monotonically and in the limit $\langle
\rho\rangle_{\rm Min}\approx 0$  it reaches the bound given in
Eq.(\ref{B_1}) and so is always strictly bounded away from 1.

\section{Conclusions}

We have examined the ratio of the quasi-local mass
to the circumferential radius, $2m/R$, for physically reasonable
spherically symmetric isolated static configurations in general relativity.
We have demonstrated how the theory always
places an upper bound on this ratio  
which lies strictly below the value it assumes when a horizon forms. 
This extends considerably earlier work on this question.

The bounds we have derived do not take into account the stability of 
these static equilibria. It would be interesting to 
know if and how these bounds get tightened when only stable configurations 
are considered.

In \cite{morris} Morris and Thorne addressed the problem of 
constructing a static wormhole. 
If one has a spherical static wormhole one must have a minimal
surface and thus $\rho + S_{\cal L} < 0$ somewhere.
This raises an interesting question. Assume one has a spherical static
wormhole and assume $\rho \ge 0$. How much `exotic material' 
violating the strong energy condition does one need?

\section*{Acknowledgements}

\noindent{We gratefully acknowledge support
from CONACyT Grant 211085-5-0118PE
to JG and Forbairt Grant SC/96/750 to N\'OM}.


\begin{references}
%
\bibitem{OV} J.R. Oppenheimer and G. Volkov {\it Phys. Rev.} {\bf 55}, 374
(1939)
%
\bibitem{Buch} H.A. Buchdahl, {\it Phys. Rev.} {\bf 116}, 1027 (1959)
%
\bibitem{Bondi} H. Bondi, {\it Proc. R. Soc.} {\bf A282} 303 (1964);
in {\it Lectures on General Relativity}
 Brandeis Summer Institute in Theoretical Physics, Volume one
(Prentice Hall, New Jersey, 1964)
%
\bibitem{Wald} R. Wald, {\it General Relativity}
(Chigaco University Press, 1984)
%
\bibitem{Wein} S. Weinberg, {\it Gravitation and Cosmology}
(Wiley, New York, 1972)
%
\bibitem{VS} A. Vilenkin and E.P.S. Shellard, {\it Cosmic Strings and
Other Topological Defects} (Cambridge Universtity Press, 1994)
%
\bibitem{Hawking}
S. Hawking and G. Ellis, {\it The large scale structure of space-time}
(Cambridge University Press, 1973)
%
\bibitem{morris} M. Morris and K. Thorne, {\it Am. J. Phys.} {\bf 56}, 385
(1988)
%

\bibitem{john} J. Friedman, K. Schleich, D. Witt, {\it Phys. Rev. Lett.}
{\bf 71}, 1486 (1993).
%
\bibitem{MTW} C. Misner, K. Thorne and J.A. Wheeler,
{\it Gravitation} (Freeman, San Francisco, 1972)
%
\bibitem{York}  Y. Choquet-Bruhat and J. York, in
{\it General Relativity and Gravitation}, edited by
A. Held (Plenum, New York, 1980)
%
\bibitem{NOM} N. \'O Murchadha, {\it Physics Letters} {\bf 60A}, 177 (1977).
%
\bibitem{murray} M. Cantor, {\it J. Math. Phys.} {\bf 20}, 1741 (1979).
%
\bibitem{Masood} A.K.M. Masood-ul-alam, {\it Class. Quantum Grav.} {\bf 5},
409 (1988)
%
\bibitem{Beig} R. Beig and W. Simon,
{\it Lett. Math. Phys.} {\bf 21}, 245 (1991)
%
\bibitem{Lind} L. Lindblom, 
{\it Phil. Trans. R. Soc. Lond.} {\bf A340}, 353 (1992)
%
\bibitem{I} J. Guven and N. \'O Murchadha,
{\it Phys Rev} {\bf D52} 758 (1995)
An extensive list of references is provided here.
%
\bibitem{II} J. Guven and N. \'O Murchadha,
{\it Phys Rev} {\bf D52} 776 (1995)

%
\bibitem{Rend} T.W. Baumgarte and A.D. Rendall,
{\it Class. Quantum Grav.} {\bf 10} 327 (1993)


%



\end{references}
\end{document}